# Lithography-free plasmonic color printing with femtosecond laser on semicontinuous silver films

Sarah N. Chowdhury[1*], Piotr Nyga[1,2*], Zhaxylyk A. Kudyshev[1], Esteban Garcia Bravo[3],

Alexei S. Lagutchev[1], Alexander V. Kildishev[1], Vladimir M. Shalaev[1],

and Alexandra Boltasseva[1]

1. School of Electrical and Computer Engineering and Birck Nanotechnology Center, Purdue University, 1205 W State St, West Lafayette IN, 47907, USA

2. Institute of Optoelectronics, Military University of Technology, 2 Kaliskiego St, Warsaw, 00-908, Poland

3. Department of Computer Graphics Technology, Purdue University, 401 N. Grant St, Knoy Hall, West Lafayette, IN 47907, USA

* equal contribution

ABSTRACT

Plasmonic color printing with semicontinuous metal films is a lithography-free and environment-friendly method for generating non-fading bright colors. Such films are comprised of islands – metal nanoparticles and their clusters of various dimensions, which resonate at different wavelengths upon external light illumination depending on the size and shape of the local particle structures. To experimentally realize systems that were optimized for achieving a broad color range and increased stability, various silver semicontinuous metal films were deposited on a metallic (Ag) mirror with a sub-wavelength-thick dielectric ($SiO_2$) spacer. Femtosecond laser post-processing was then introduced to extend the color gamut through spectrally induced changes from blue to green, red, and yellow. Long-term stability and durability of the structures were addressed to enable non-fading colors with an optimized overcoating dielectric layer. The thickness of the proposed design is on the order of 100 nanometers, and it can be deposited on any substrate. These structures generate a broad range of long-lasting colors in reflection that can be applied to real-life artistic or technological applications with a spatial resolution on the order of 0.3 mm or less.

INTRODUCTION

We currently have about sixteen million colors available in the realm of natural and artificial palette[1], and still, everyday people are yearning for new shades of color. However, to achieve the full-range color palette, synthetic (artificial) dyes are commonly used, which often expel toxic chemicals. Such dyes are often not non-biodegradable, carcinogenic[2], can alter the physical and chemical properties of soil, deteriorate water bodies, and cause irreversible harm to flora and fauna[3]. In turn, non-toxic and environmentally-friendly coloration using plasmonic nanostructures has been in vogue for centuries[4,5], and has recently been used for sub-wavelength printing[6–20]. Even though exemplary work on color generation using plasmonic structures has been done with sub-wavelength resolution e-beam lithography[8,14,16–18,20,21] and ion milling[7,13,15,19], the high fabrication cost has so far hindered the scalable reproduction of plasmonic colors for coatings, fine arts and technological applications.

Semicontinuous metal films (SMF), which are disordered, island-containing films made of metallic nanoparticles and their clusters of various sizes and shapes, have recently been explored for plasmonic coloration[22–24]. Such films can be readily achieved in a simple deposition process where a metallic layer is deposited on a planar substrate close to the percolation threshold, which means that instead of forming a continuous film, disconnected metal islands of different sizes and shapes are formed. During metal deposition on a dielectric substrate, one initially has isolated metal particles on the substrate; with an increase in the metal filing factor, the particles first form fractal clusters of various sizes and shapes and then, at the percolation threshold, a continuous metal path ("infinite cluster") is formed, which connects the different ends of the film so that the DC current can pass ("percolate") through the system[25,26]. Such random metal-dielectric films near the percolation threshold are often referred to as semicontinuous metal films. It was predicted by Shalaev and later shown experimentally that plasmon modes in fractals and percolation films are localized, forming the "hot spots", where the local fields are significantly enhanced compared to the incident light, resulting in strongly enhanced nonlinear responses and many other interesting phenomena[25–33]. The difference in local structures of the areas where plasmon modes are localized, resulting in the hot spots, means that they resonate at different frequencies (and different polarizations). A large variety in such local structures in turn implies that the localized plasmons supported by these structures cover all together an extremely broad spectral range, from UV to far-IR, which characterizes the absorption and reflectance of SMFs[25,26,34]. At sufficiently high intensities, the strongly enhanced local fields in the hot spots can cause the local heating of the resonating nanostructures and, as a result, locally modify their shapes and geometries via selective melting, sintering, or fragmentation so that these structures do not resonate any more at the given frequency and polarization[35–43]. As a result of such local photomodification, spectrally and polarization selective changes occur in the transmittance, reflectance and absorption spectra due to these local structural changes, occurring in the nanometer-scale areas of a SMF[35–43]. Such spatially, frequency and polarization selective photomodification enables a gradual change in the optical spectra, which in turn allows for a highly controllable change in colors. This inexpensive, simple, and environmentally-friendly method of color generation has been recently demonstrated with single-layer SMFs and with SMFs deposited over a dielectric spacer on top of a metallic backplane mirror[44] to show promising results for macroscopic printing applications[22–24,43,44]. Laser printing on other plasmonic random[6,45–50] and ordered[51] structures have also been recently reported.

In this paper, we report studies of plasmonic colors in so-called gap plasmon structures[52–56]. Such geometry is created of an optically thick silver film (a mirror) deposited on a glass substrate, followed by a thin dielectric (silica, $SiO_2$) spacer and a silver SMF. We refer to such structures as the semicontinuous metal films on the mirror (SMF/M). We employ the dielectric spacer and mirror to utilize the gap-plasmon mode–coupling of the SMF's nanoparticles with the metallic mirror[57]. While we have earlier proposed these structures[24,58–60], neither a thickness dependence nor optimization of colors has yet been studied. In this paper, we aim at achieving the overall optimal design that generates a wider gamut of non-fading vibrant colors and address the crucial problem of the chemical stability of Ag SMFs using a dielectric coating. We explore coatings fabricated with both conventional electron-beam evaporation and atomic layer deposition (ALD). The developed fabrication technique is lithography-free; it overcomes the core disadvantages of expensive electron-beam lithography for the generation of colors. Due to a very thin overcoating layer, the thickness of the entire structure is on the order of one hundred nanometers (~150 nm). It can be deposited on any substrate that can sustain the thermal heating due to laser illumination under fluences ranging from 5 mJ/cm$^2$ to 400 mJ/cm$^2$. The fabrication approach requires less quantity of chemical elements for the overall process compared to other macroscopic printing techniques. Furthermore, it overcomes the drawback of a lower resolution in lithography-free color generation techniques with a beam size of 0.3 mm that can be decreased even further based upon the purpose of a given application. Finally, we present several fine-art pieces, using our in-house laser photo-modification setup, laser-printed on the SMF/Ms and protected with the ultra-thin dielectric coating. The paper has been organized as follows. First, we overview the proposed SMF/M structure and its optimization in Section 1. The optimized structures post-processed with laser photomodification for a wider gamut of colors are discussed in Section 2. The durability of the proposed structures with stable colors is addressed in Section 3. Finally, fade-free artistic images are presented in Section 4 as select application examples.

**RESULTS AND DISCUSSION**

**1. Dependence of the SMF/M color on the thickness of the top Ag SMF layer**

Semicontinuous metal films are made of disordered nanoparticles of random size, shape, and position distributed over a substrate. The morphology of SMFs and metal filing factor strongly depend on the amount of metal mass deposited and details of the deposition process. Thus, the amount of the deposited mass can drastically alter the optical properties of the film depending on how far/close one is to the percolation threshold, leading to a diverse color palette. Unfortunately, for a given metal, the possible colors from a single-layer SMF are somewhat limited.[24] Another approach to changing the optical spectra is through utilizing gap plasmon resonance between two metal layers separated by a relatively thin dielectric layer[23,24,57,61]. Such structures have been studied in our previous study, where gap plasmon resonances between SMF nanostructures and a metallic mirror layer separated by a silica layer were utilized. The schematic view of the structure is shown in Figure 1. The structure comprised of an Ag SMF layer of thickness, $t_{Ag} = 10$ nm with an overcoating layer of silica of thickness, $t_c = 30$ nm. Although that structure[24] showed stable colors, a sufficiently wide gamut was not achieved.

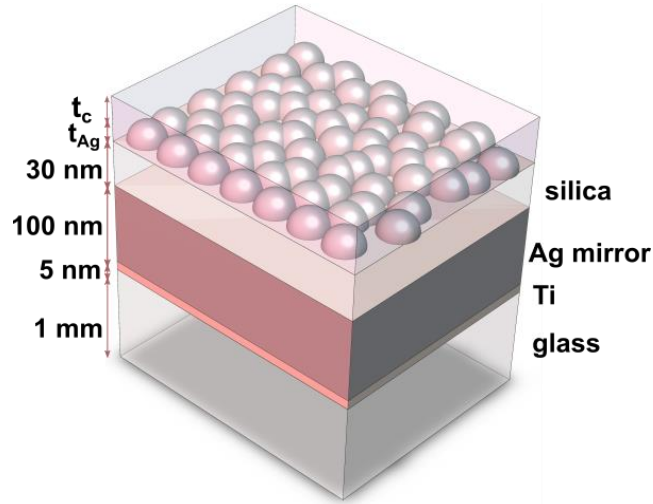

Figure 1. Schematic view of the SMF/M structure. The thickness of the top $A_g$ SMF layer is $t_{Ag}$ and the thickness of the overcoating layer is $t_c$, the thickness of the substrate is scaled to fit the figure.

To achieve a broader range of stable colors, we first changed the Ag SMF layer to explore the correspondence between the deposited metal nanoparticles and the resulting optical spectra. From our earlier studies it has been known[24] that, for the SMF/M type structures, thickness of above 17 nm ($t_{Ag} \geq 17$ nm) results in Ag films above percolation with a metallic optical response, showing substantial reflectance. Moreover, upon the photomodification, the laser-induced heat in the hot spots in such film, instead of being absorbed locally, is dissipated over the film, due to the numerous continuous metallic paths facilitating the heat transfer. The SMF/Ms of that class also show low spectral and spatial selectivity when modified. Hence, we analyzed the structures below percolation, with $t_{Ag} < 17$ nm, for which the incident energy is absorbed locally, mainly close to the hot spots around the Ag nanoparticles. In this case, there is limited heat transfer from the areas of the hot spot as the film below percolation is made of isolated nanostructures.

Figure 2a shows the reflectance spectra for the Ag SMF with thicknesses ranging from $t_{Ag} = 5$ to 17 nm. Corresponding visible colors are also presented along with the SEM images that show the relationship between the film morphology and the optical spectra. We note that, with the increasing thickness, the size of the metal nanoparticles changes, corresponding to the change in the absorption and the reflected color. Initially, we have been unable to achieve robust SMFs with $t_{Ag} \leq 5$ nm, due to the inherent oxidizing nature of Ag[62], and those SMF/Ms began to lose their initial color quickly (in a matter of days). Hence, we protect the structures with a silica layer of thickness $t_c = 30$ nm. The silica-protected structures demonstrated high stability compared to the uncoated films. Moreover, the protective silica coating shifts the resonances of the SMF nanostructures to longer wavelengths, producing the reflected colors in the blue region (Figure 2b). The corresponding CIE 1931 color diagram of uncoated and coated SMF/Ms is shown in Supplementary Information (Figure S1). These initial colors of the coated SMF/M structures open new possibilities of obtaining different colors through post-laser modification for tunable spectral selectivity.

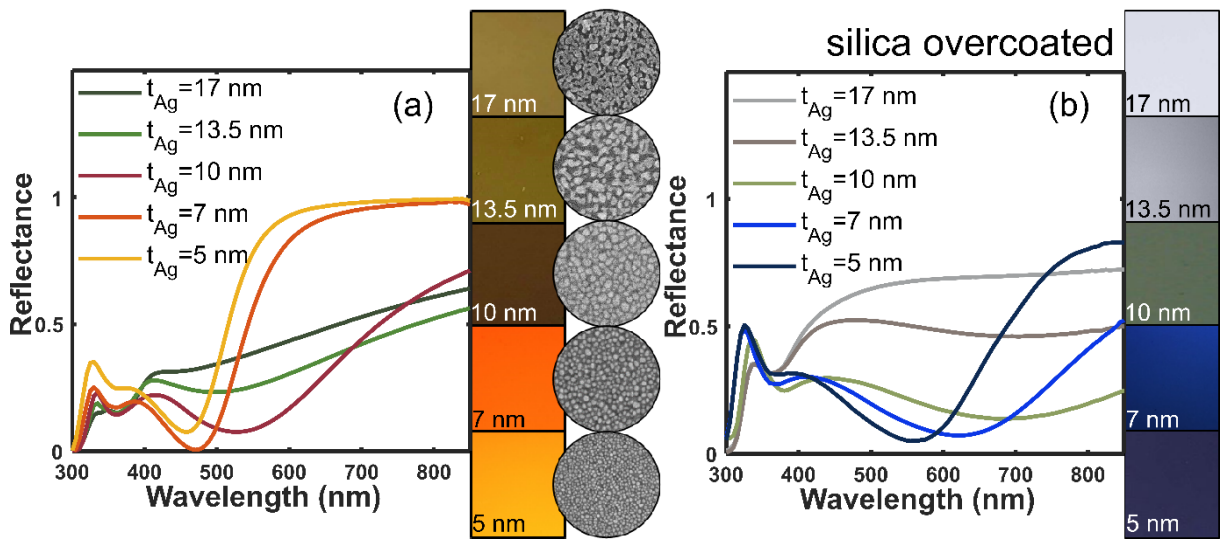

Figure 2. Reflectance spectra of Ag SMF/M structures with different Ag thicknesses ($t_{Ag}$) along with their corresponding colors for (a) without and (b) with the 30 nm silica overcoating layer. The color squares are the SMF/M surface images captured using an optical microscope with 10x objective lens (Nikon Eclipse, L150). Corresponding SEM images are also inserted in parallel which show the change in the film structure as the Ag thickness increases. Diameter of the SEM circled image is 0.4 μm.

2. **Laser color printing on SMF/M**

For inducing spectrally selective changes in the optical spectra, the fabricated SMF/M structures were subjected to laser post-processing that induced local heating of the resonating nanostructures[63]. For $t_{Ag} > 10$ nm, the SMF/M structures are quite reflective (Fig. 2), and hence, laser post-processing was not able to induce changes in the absorption spectra concluding to modifications on samples with $t_{Ag} \leq 10$ nm.

For laser post-processing, we use linearly polarized femtosecond laser pulses (repetition rate 1 kHz, pulse duration 80 fs, wavelength 800 nm). The results of the laser modification on the SMF/M sample with Ag thickness, $t_{Ag} = 7$ nm are presented in Figure 3 (while the results for $t_{Ag} = 10$ and 5 nm are presented in Supplementary Information, Figures S2 and S3). The laser fluence is gradually increased with the above-mentioned laser parameters from 5 mJ/cm² to the value of damage threshold (on the order of 200 mJ/cm²). The corresponding reflectance spectra was then measured using a spectrometer equipped with an integrating sphere (see Fig. 3a, Fig. S2a, and Fig. S3a). After photomodification, a "step function" reflectance response in the visible range is observed, which moves to shorter wavelengths as the laser fluence increases. Such behavior can be attributed to the gradual fragmentation of nanostructures in Ag SMF and formation of spherical nanoparticles[41]. Thus, the SMF/M for the longer wavelengths (SI, Fig. S4) behaves as a metallic mirror protected with a dielectric layer. At the same time at low fluences, there is an increase of absorption at shorter wavelengths due to the formation of new nanostructures absorbing at shorter wavelengths. Photoinduced damage of the bottom silver mirror occurs at the fluences above 200 mJ/cm². The induced changes in the optical spectra then result in a wide range of reflected colors. We observe a broader color gamut from the CIE 1931 color map, which has been generated from the corresponding reflectance spectra from $t_{Ag} = 7$ nm in Fig. 3b. We obtain colors from blue to yellow, whereas earlier[24] the gamut was limited to orange, and the colors for shorter wavelengths were also not achieved (Supplementary Information, Fig. S5). From the CIE color maps for each case (Fig. 3b and Supplementary Information, Fig. S2b and Fig. S3b), we observed that, the Ag SMF/M with thickness, $t_{Ag}$ = 7 nm gives the widest range from blue to green, red, and yellow.

Although the initial tests for the silica protective coating ($t_c$ = 30 nm) conducted on the Ag SMF/M sample with $t_{Ag}$ = 10 nm ensured chemical stability[24], such a protective layer appears to be inadequate for lower Ag SMF thicknesses ($t_{Ag}$ = 5 and 7 nm) to abate oxidation/sulfidation completely, as those structures degrade over time. Hence, it is necessary to additionally optimize the protective layer to achieve chemical stability and color vibrancy in SMF/Ms with $t_{Ag}$ < 10 nm.

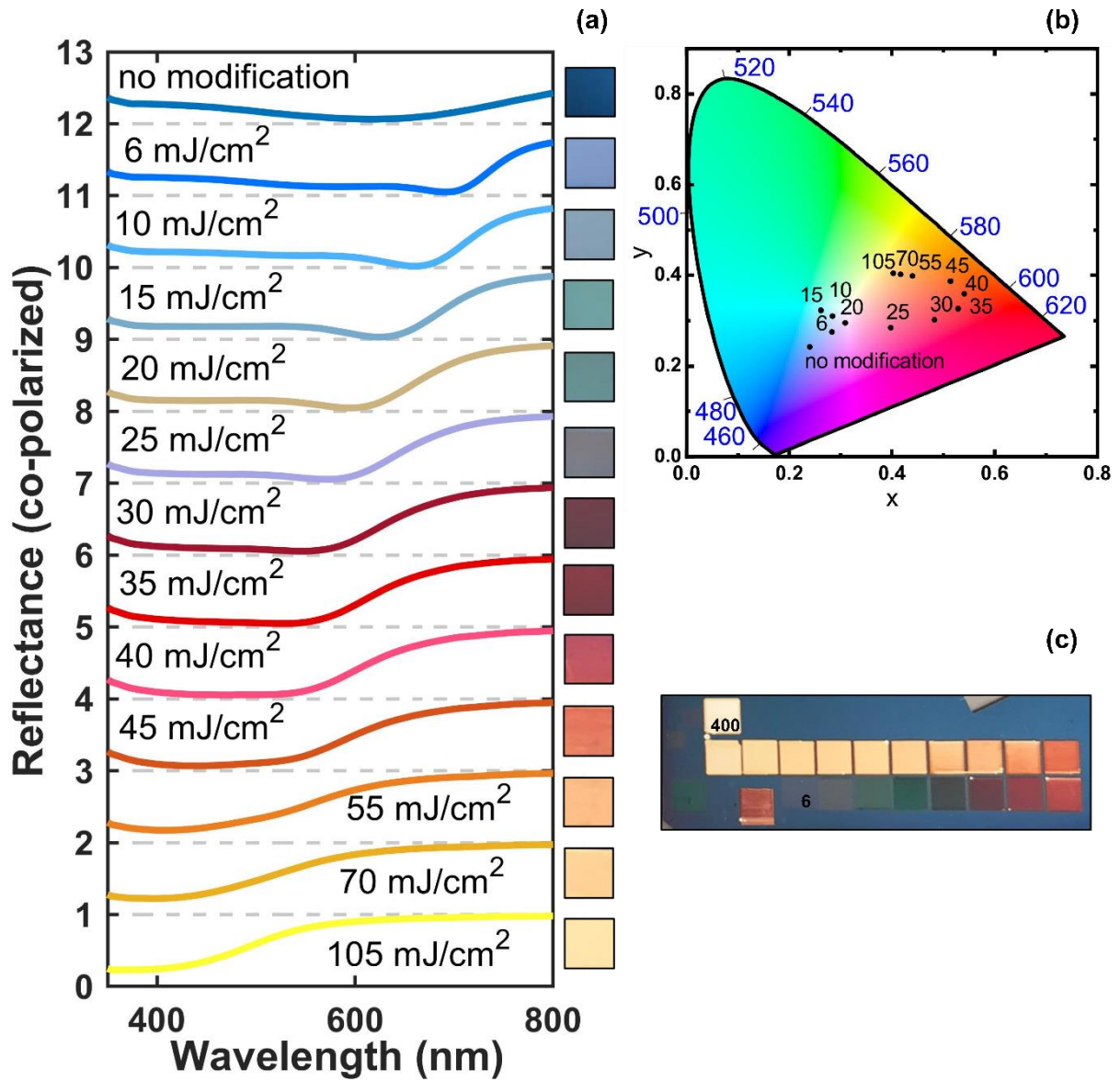

Figure 3. Laser modification of Ag SMF/M ($t_{Ag}$ = 7 nm) with a silica protective coating, $t_c$ = 30 nm. (a) Reflectance spectra and (b) corresponding CIE 1931 color diagram of areas photomodified with different laser fluences (cross-polarized in Supplementary Information, Fig. S6). The inset squares represent the generated color palette recorded using unpolarized light. (c) Optical image of photomodified squares from 6 mJ/cm² (marked '6') to 400 mJ/cm² (marked '400').

## 3. SMF/M stability studies

To address the issue of chemical stability, one approach is to further increase the thickness of the protective silica coating to sustain chemical damage. We initially performed numerical simulations with a commercially available 3D finite-difference time-domain (FDTD) solver (Lumerical Inc., FDTD Solutions)[64]. To precisely retrieve the optical response of the random

Ag SMF films, the binarized SEM images of the fabricated film have been imported into the simulation domain using a build-in Lumerical material import parser. We extracted the relative permittivities of Ag and silica through spectroscopic scan using a J. A. Woollam V-Vase UV-vis-NIR spectroscopic ellipsometer. The optical response of the SMF type films depends on the distribution of the plasmonic particles of different sizes. Hence, the response of the structure should be statistically averaged over a sufficiently large number of realizations. Thus, we select 25 distinct regions with an area of 500 × 500 nm$^2$ at various spots of the same sample. The overall response of the SMF film is assessed by averaging the power transmission/reflection coefficients over these realizations. This simulation approach is similar to our previous studies on semicontinuous metal films[24,42,65], except for the use of the commercial software and better automation of the image processing and geometry parsing process.

The simulations are performed for the Ag SMF/M structure with $t_{Ag}$ = 10 nm with a gradual increase in the thickness of the overcoating silica layer from $t_c$ = 30 nm to $t_c$ = 90 nm and then experimentally tested (SI, Fig. S7a and Fig. S7b). The experimental data corroborated our simulations. Also, the thick silica overcoating layer ($t_c$ = 90 nm) showed a redshift in the resonance dip and a broadening of spectrum similar to the simulation results (SI, Fig. S7). Unfortunately, with a thick overcoating layer, laser modification produces lower saturation and hue of colors (Supplementary Information, Fig. S7c and Fig. S7d). Although we tested the chemical stability of the thick silica overcoating layer ($t_c$ = 90 nm) by measuring the reflectance spectrum of the freshly fabricated sample and remeasured after eleven months (Supplementary Information, Fig. S8), such a low vibrancy has been a critical impediment towards using a more robust thick silica overcoating layer. Moreover, an overall structure that is relatively thin is also desired for scalable production with less materials.

Another approach is to change the bottom-up coating technique for passivation, which might not seal the three-dimensional Ag nanoparticles of top SMF properly, leaving areas shadowed and less immune to chemical reactions. Atomic layer deposition is an effective method of passivation of nanostructures that yields conformal thin films[66,67] at low temperatures. Our Ag sample is also susceptible to temperature variance, and ALD provides an additional benefit of thin layer deposition at low temperatures[68].

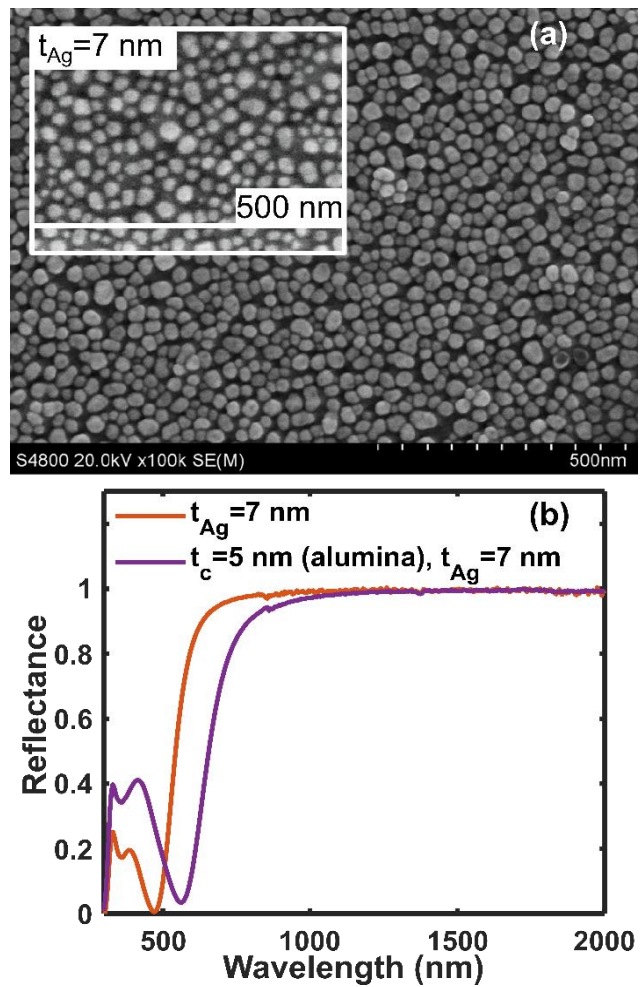

Figure 4. (a) SEM image of the Ag SMF/M ($t_{Ag}$ = 7 nm) with an alumina protective coating, $t_c$ = 5 nm. The inset shows a SEM image of uncoated Ag SMF/M ($t_{Ag}$ = 7 nm) (b) Reflectance spectra of Ag SMF/M ($t_{Ag}$ = 7 nm) with and without an alumina protective coating.

We tested passivation of the SMF/M surface with a thin coating ($t_c$ = 5 nm) of alumina ($Al_2O_3$), using ALD from TMA and water precursors (Fig. 4a). The alumina coating redshifts the optical spectra (Fig. 4b) as expected due to a higher refractive index of alumina as compared to air. Before performing laser modifications on the alumina-coated SMF/M structures, we corroborated their stability. We compared the reflectance spectra of Ag SMF/M ($t_{Ag}$ = 7 nm) coated with $t_c$ = 30 nm (silica), which also gave vibrant colors, to Ag SMF/M ($t_{Ag}$ = 7 nm) with the alumina coating, $t_c$ = 5 nm (Fig. 5). The reflectance of those structures was measured (i) immediately after fabrication and (ii) after several months of storage in the ambient atmosphere. While we observe significant changes in the spectra (mainly shift of reflectance minimum to shorter wavelengths) for the SMF/M coated silica, negligible changes are observed for the alumina-coated SMF/M.

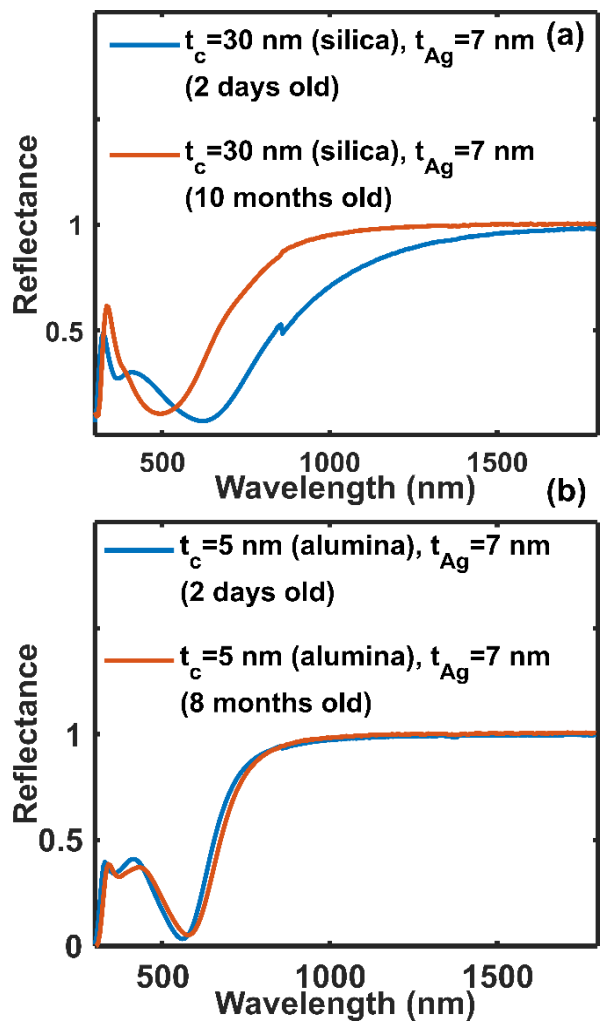

Figure. 5. Reflectance spectra of (a) Ag SMF/M with $t_{Ag}$ = 7 nm overcoated with $t_c$ = 30 nm (silica) and (b) Ag SMF/M with $t_{Ag}$ = 7 nm overcoated with $t_c$ = 5 nm (alumina).

As the alumina-coated SMF/Ms demonstrated higher stability, we further explored the possibility of their laser modification for the generation of the broad color range. We followed the same procedure as for the silica-coated samples. Figure 6a shows the reflectance spectra of the Ag SMF/M with $t_{Ag}$ = 7 nm overcoated with $t_c$ = 5 nm (alumina) photomodified with various laser fluences from 10 to 250 mJ/cm$^2$. We notice a similar trend of change in the reflectance spectra as the laser fluence is increased. At higher fluence (above 150 mJ/cm$^2$), a saturated hue in the yellow region is observed (Fig. 6b) and a further increase does not generate significant change in the color, but gradual damage of the bottom mirror occurs (above 250 mJ/cm$^2$). For the Ag SMF/M with $t_{Ag}$ = 7 nm overcoated with $t_c$ = 5 nm (alumina) there is a gradual transformation to spherical nanoparticles with increasing energy density (Fig. 7). The Ag nanoparticles seemed to deform and detach from the bottom layer, while the alumina coated protective layer keeps the particles intact, showing the original extrusions of the particle shape[67]. We also checked the stability of the reflectance of a photomodified area and compared the results (SI, Fig. S9). In all cases, the alumina coated structures demonstrate superior stability as compared to $t_c$ = 30 nm (silica).

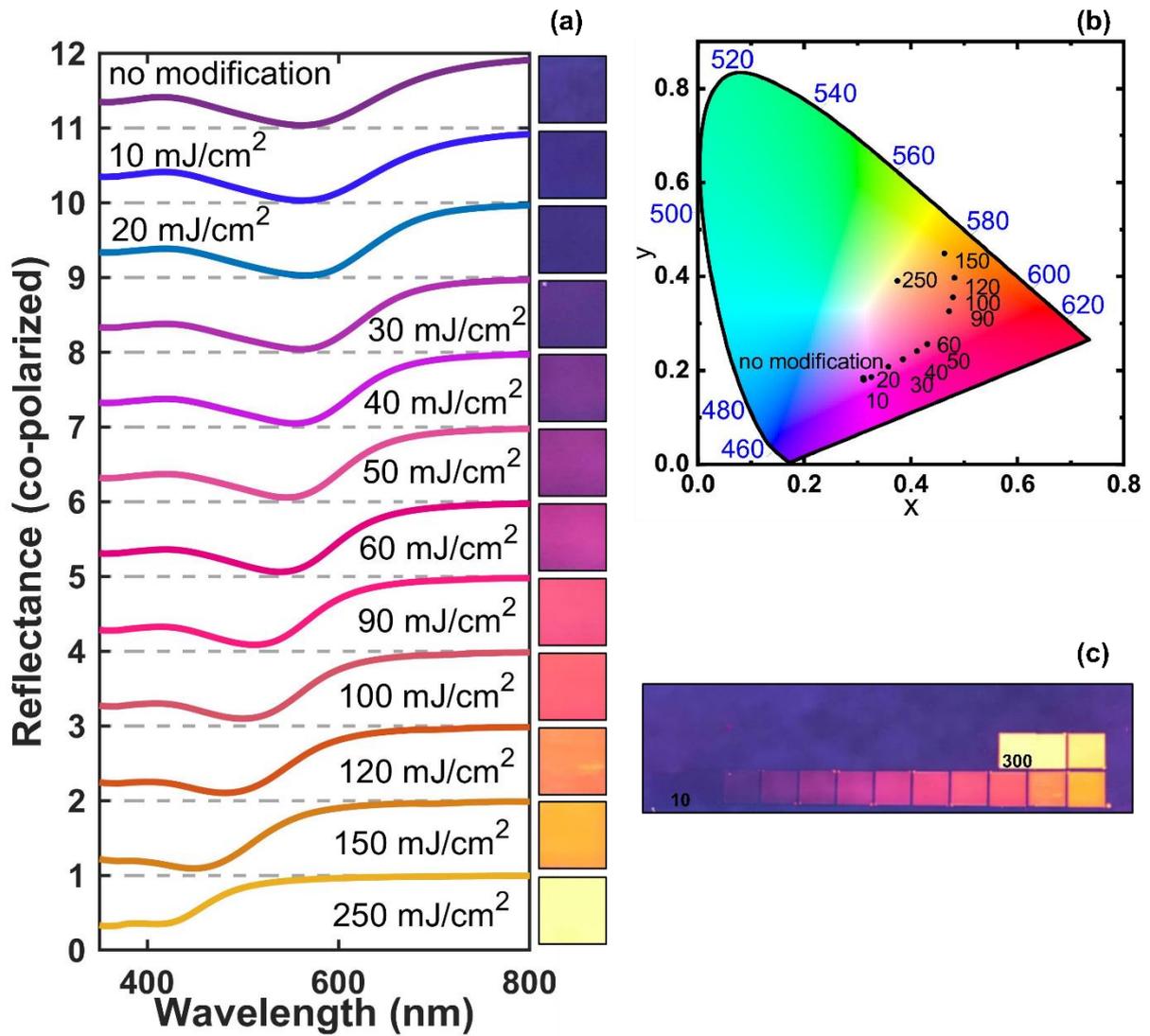

Figure 6. Laser modification of Ag SMF/M with $t_{Ag}$ = 7 nm overcoated with $t_c$ = 5 nm (alumina). (a) Reflectance spectra and (b) corresponding CIE 1931 color diagram of areas photomodified with different laser fluences. The inset squares represent the generated color palette recorded using unpolarized light. (c) Optical image of the photomodified squares from 10 mJ/cm$^2$ (marked '10') to 300 mJ/cm$^2$ (marked '300').

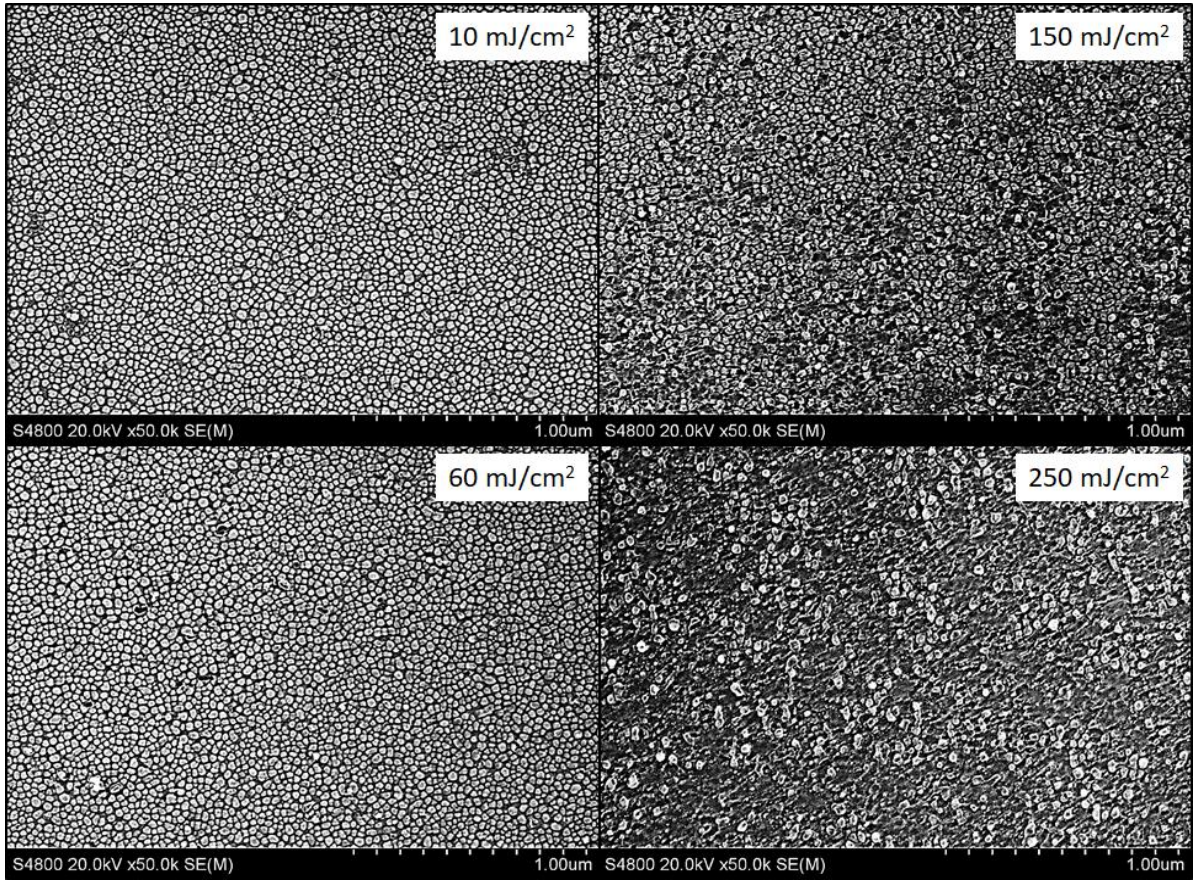

Figure 7. SEM images of Ag SMF/M with $t_{Ag}$ = 7nm overcoated with $t_c$ = 5 nm (alumina) photomodified with different energy densities.

### 4. Fine-art application

We explored the visual capabilities of the proposed approach in terms of quality, resolution, and aesthetics of the laser modification method. We reproduced different fine designs (Fig. 8) with the in-home built laser scanning setup. We built an optical arrangement for the Purdue 2050: Conference of the Future "Engineering for Arts" exhibit (Figs. 8a and 8b) that project reflections at a bigger scale (24"×24"). The setup showcased different Mandala designs and other artifacts (Fig. 8). The proposed strictures and the laser setup can be applied to print arbitrary designs on rigid and flexible substrates that opens the potential for both research and practical applications.

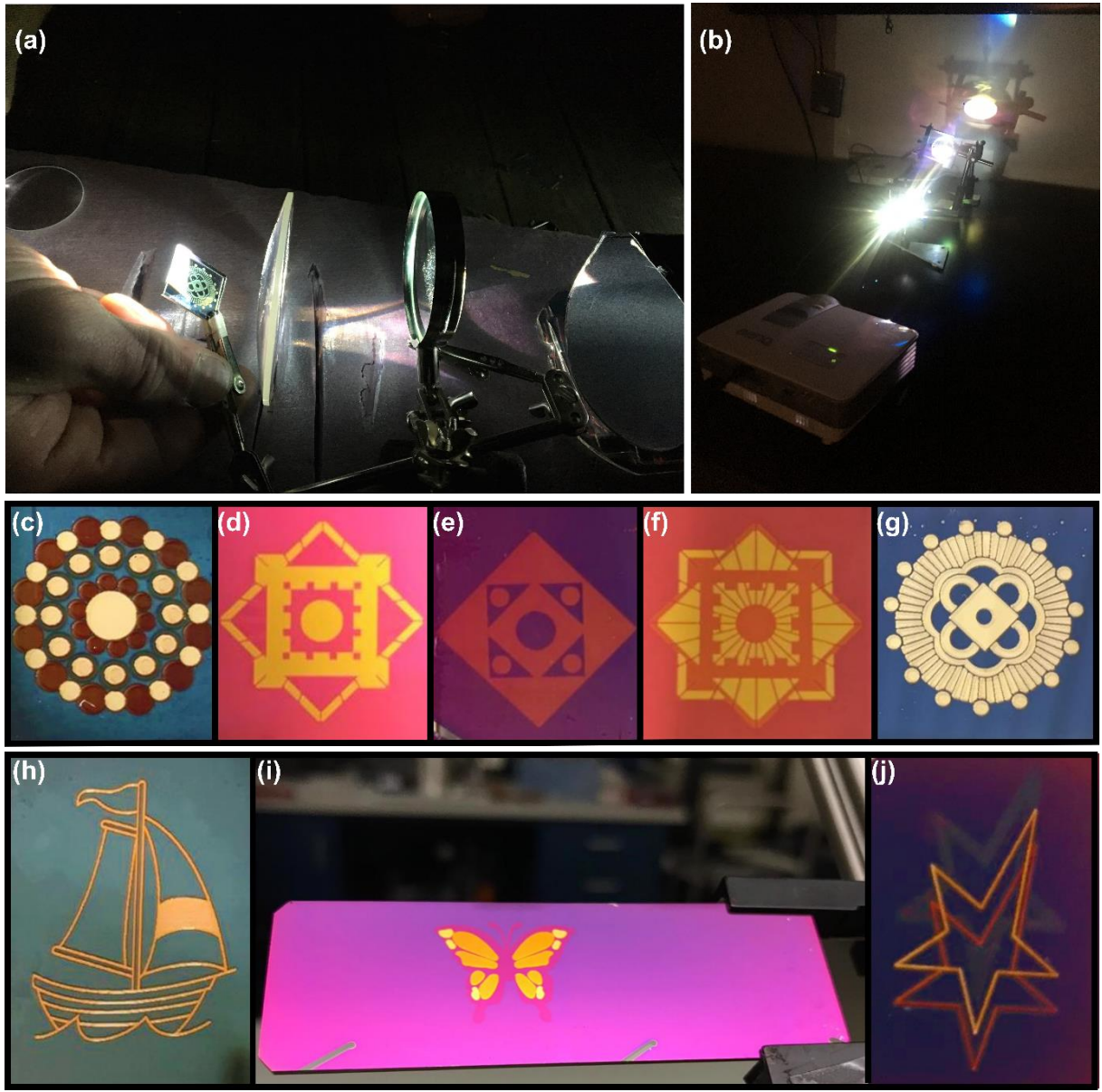

Figure 8. (a-b) Optical setup for projecting reflections from the surface of our fabricated sample at a bigger scale (24"×24"). (c-g) Different color images printed on Ag SMF/Ms. The designs (c-g) were produced as part of the exhibit titled 'Plasmonic Mandalas' in Purdue 2050: Conference on the Future ([The Exhibit](#)). A demonstration of the magnified dimension has been recorded here ([Plasmonic Mandalas](#)).

## CONCLUSION

Random metal nanoparticles and their clusters of various sizes and shapes comprising semicontinuous metals films (SMFs) can provide a wide variety of colors. In this paper, we demonstrated that SMFs with a dielectric spacer on a mirror (SMF/M) could be designed for macroscopic color applications through simple variation of the silver layer thickness. The chemical stability of Ag films is achieved with protective coatings of either a thick silica layer or a thin layer of ALD coated alumina and validated through exposure to air. Laser illumination of SMF/Ms with different energy densities gave rise to a broad range of colors through thermally-induced changes in the shape and size of Ag nanoparticles. We also showed the long-term stability of the laser-modified SMF/Ms and designed a broad, robust, and non-fading color gamut. The proposed SMF/M structures have the potential for providing even broader range of colors by utilizing different materials and geometries. A more rigorous study can be conducted on the effect of various surrounding media on the localized thermal heating of plasmonic nanoparticles, for example, through thermal reflectance measurements. Laser post-processing can further be optimized by varying line spacing, scanning speed, exposure time, beam size, and other laser parameters to enhance the overall color palette. The develop setup can be used for more advanced artistic designs and technological applications that require higher resolution.

**METHODS**

**Fabrication of SMF/M structure**s. The SMF/M structures formed from a silver SMF, silica spacer, and a silver mirror deposited on a glass substrate were fabricated in a single process using an electron-beam physical vapor deposition (PVD) technique. The glass substrates were pre-cleaned with Piranha solution (3 parts $H_2SO_4$:1 part $H_2O_2$) for 15 minutes and thoroughly rinsed with distilled water. After drying out with nitrogen gas, the substrates were then sonicated in solvents (toluene, acetone, and isopropyl alcohol) and dried thoroughly[42]. Next, a titanium adhesion layer , silver mirror, , silica spacer and Ag SMF were deposited in a high-vacuum deposition chamber (PVD Products, Inc., base pressure $2.6\times10^{-6}$ torr) at room temperature. Silicon dioxide ($SiO_2$, 99.99% purity), titanium (Ti, 99.99% purity), and silver (Ag, 99.99% purity) from the Kurt J. Lesker Company were used for fabricating all structures. The deposition rate (1 Å/s for all materials) and layer thickness were monitored with a quartz crystal microbalance. The protecting layer on top of the structure was deposited in two ways: electron beam evaporation and atomic layer deposition. The silica overcoating layer was deposited within the same process in the PVD chamber. The atomic layer deposition was used to deposit a thin protective layer of alumina at 100 °C. In the ALD process, trimethylaluminum (TMA, $Al(CH_3)_3$) and water ($H_2O$) were alternately entrained with an argon carrier flow using gas switching valves.

The process of forming a monolayer of alumina started with pumping TMA pulse for 0.06 s. At this step, TMA reacted with the hydroxylated surface of SMF/M after exposure to air to form -OH bonds resulting in a monolayer. The unreacted molecules of TMA were then purged with argon carrier gas for 10 s. Water with a pulse of 0.06 s is afterward pulsed. That process removed the $CH_3$ groups, creates Al-O-Al bridges, so that a passivated-surface with Al-OH is formed. $CH_4$ (methane) was formed as a gaseous byproduct at the end of this step. The unreacted $H_2O$ and $CH_4$ molecules were purged for 10s. This step ended a single cycle. One cycle resulted in the deposition of approximately 1 Å of alumina.

**Laser post-processing and design setup.** Laser processing on the SMF/Ms was performed in an ambient atmosphere using an ultrafast Ti:Sapphire femtosecond laser (Spectra-Physics, Solstice Ace;1 kHz, 80 fs, 800 nm, linear polarization). The laser beam was focused using a single lens and the $1/e^2$ Gaussian beam size was determined using the knife-edge technique. To print areas of uniform color, samples were mounted on a motorized XYZ stage (Zaber Technologies Inc.) capable of raster scanning and controlled with a computer interface. The scanning speed was maintained at 3 mm/s. To ensure uniformity of modification over the large area, we use 50 μm Y-axis (raster) step. A Matlab generated code for the Zaber XYZ stage control toolbox was used to print different designs on the samples. We used an optical microscope (Nikon Eclipse, L150) as well as a general camera to capture the color images of printed structures.

**Sample characterization.** A field emission scanning electron microscope (FESEM, Hitachi S-4800) was used to characterize the nanostructure of uncoated SMFs. The semicontinuous metal films of SMF/M overcoated with 30 nm and 90 nm silica layer were not characterized with SEM; it was not possible to visualize the metallic nanostructures through a relatively thick silica layer. In contrast, the SEM images for the case of the SMF/Ms with $Al_2O_3$ coatings were recorded, as these protective layers were significantly thinner. Total transmittance and total reflectance spectra of as-fabricated and photomodified structures were measured using a spectrophotometer (Perkin Elmer, Lambda 950) equipped with an integrating sphere (150 mm) module (8 deg. angle of incidence used for reflectance measurement) and polarizers. Spectralon was used as a reference sample for reflectance measurements. The calculation of CIE 1931 color coordinates from reflectance spectra was performed with ORIGIN™ software[69,70].

Supporting Information 1 Available: This material is available free of charge via the Internet at http://pubs.acs.org.

- CIE 1931 color diagram of variable Ag thickness of semicontinuous metal film with mirror (Ag) and dielectric spacer ($SiO_2$) (SMF/M) with and without the overcoating $SiO_2$ layer.
- Laser modification of Ag SMF/M with $t_{Ag} = 10$ nm overcoated with $t_c = 30$ nm (silica)
- Laser modification of Ag SMF/M with $t_{Ag} = 5$ nm overcoated with $t_c = 30$ nm (silica)

- Reflectance spectra (extended wavelength range) of Ag SMF/M with $t_{Ag} = 7$ nm overcoated with $t_c = 30$ nm (silica)

- CIE 1931 color diagram of 30 nm overcoated 10 nm Ag SMF/M from reference[24]

- CIE 1931 color diagram of Ag SMF/M with $t_{Ag} = 7$ nm overcoated with $t_c = 30$ nm (silica) (cross-polarized reflectance measurement)

- Reflectance spectra (simulated and experimental) of variable top silica layer of Ag SMF/M with $t_{Ag} = 7$ nm for stable structures. Laser modification of Ag SMF/M with $t_{Ag} = 10$ nm overcoated with $t_c = 90$ nm (silica)

- Stability test on Ag SMF/M with $t_{Ag} = 10$ nm overcoated with $t_c = 90$ nm (silica)

- Stability test on one of the photomodified areas of Ag SMF/M with $t_{Ag} = 7$ nm overcoated with $t_c = 30$ nm (silica) and $t_c = 5$ nm (alumina)

Supporting Information 2 Available: This material is available free of charge via the Internet at http://pubs.acs.org.

- YouTube video link for the demonstration at the exhibit of Purdue 2050: Conference on the Future (https://youtu.be/J1yrmwz493Y)

AUTHOR INFORMATION

**Corresponding Author**

piotr.nyga@wat.edu.pl

2 Kaliskiego St, Warsaw, 00-908, Poland

aeb@purdue.edu

1205 W State St, West Lafayette IN, 47907, USA

**Author Contributions**

P.N. conceived the idea of the laser color printing on SMF/M structures. S.N.C. and P.N. performed fabrication and optical characterization, data analysis and wrote the initial draft. S.N.C. did the SEM characterization. E.G. and S.N.C. prepared designs of images for laser printing, Z.K. performed FDTD simulations. P.N., S.N.C. and A.S.L. performed laser modification. P.N., A.B., A.V.K., and V.M.S. supervised the project.

All authors analyzed and discussed the results and participated in the preparation of manuscript.

**Funding Sources**

Purdue team acknowledge financial support by the Air Force Office of Scientific Research grant FA9550-18-1-0002 and funding for the laser system by the Office of Naval Research award No. N00014-17-1-2910. PN acknowledges financial support by the Military University of Technology UGB 502-6700-23-759 grant and Fulbright Senior Award Scholarship awarded by the Polish-U.S. Fulbright Commission.